\documentclass[twocolumn,floatfix,showpacs,amsmath,amssymb,aps,prl]{revtex4}
\usepackage{graphicx}
\usepackage{dcolumn}
\usepackage{bm}
\usepackage{epsfig}

\def\be{\begin{equation}}       \def\ee{\end{equation}}
\def\bea{\begin{eqnarray}}      \def\eea{\end{eqnarray}}
\def\pp{\parallel}

\begin{document}
\def \cH{{\cal H }}
\def\nd{{^{\vphantom{\dagger}}}}
\def\yd{^\dagger}
\def \bea{\begin{eqnarray}}
\def \eea{\end{eqnarray}}

\title{Current Carrying Ground State in a Bi-layer Model}

\author{Sylvain Capponi}
\affiliation{Laboratoire de Physique Th\'eorique UMR 5152,
Universit\'e Paul Sabatier, 118 route de Narbonne, 31062 Toulouse,
France}
\affiliation{Department of Physics, McCullough Building, Stanford
University, Stanford  CA~~94305-4045}
\author{Congjun Wu}
\author{Shou-Cheng Zhang}
\affiliation{Department of Physics, McCullough Building, Stanford
University, Stanford  CA~~94305-4045}
\date{\today}

\begin{abstract}
Strongly interacting systems have been conjectured to
spontaneously develop current carrying ground states under certain
conditions. We conclusively demonstrate the existence of a
commensurate staggered interlayer current phase in a bi-layer
model by using the recently discovered quantum Monte-Carlo
algorithm without the sign problem. A pseudospin SU(2) algebra and
the corresponding anisotropic spin-1 Heisenberg model are
constructed to show the competition among the staggered interlayer
current, rung singlet and charge density wave phases.
\end{abstract}
\pacs{71.10.Fd,71.10.Hf, 71.30.+h, 74.20.Mn } \maketitle

Strongly correlated systems can spontaneously break symmetries of
the microscopic Hamiltonian. A particularly interesting class of
ground states spontaneously break the time reversal symmetry
and carry a persistent current in the ground state. Such states
are known by different synonyms, {\it e.g.} the orbital
antiferromagnetic phase (OAF), the staggered flux (SF) or the
D-density wave (DDW) phase. In the context of high Tc
superconductivity, these current carrying ground states have been
proposed as competing states for the pseudogap
phase\cite{AFFLECK1988,HSU1991,WEN1996,VARMA1999,CHAKRAVARTY2001A,schroeter2002}.
The SF or the DDW phase  has the attractive feature that the nodal
quasi-particles have an energy spectrum similar to that of the
$d-$wave superconducting state.

Whenever new ground states are proposed, it is important to
establish for which microscopic Hamiltonian such states are
realized. Because of their relative simplicity and availability of
reliable analytical and numerical methods, the ladder system has
been used as a theoretical laboratory to investigate the DDW
phase. Weak coupling bosonization methods combined with the
renormalization group (RG) analysis on extended two-leg Hubbard
ladders show the existence of commensurate DDW phase at
half-filling \cite{FJAERESTAD2002,WU2003A,TSUCHIIZU2002} and
incommensurate power law fluctuating DDW order away from
half-filling\cite{SCHULZ1996,ORIGNAC1997,WU2003A}. While the DDW
state does not appear to be the ground state of the t-J ladder~\cite{SCALAPINO2001,TSUTSUI2002},
numerical works using the density matrix renormalization
group(DMRG) found commensurate DDW order at
half-filling\cite{MARSTON2002} and incommensurate DDW order at low
doping\cite{SCHOLLWOCK2003} in a ladder model first proposed by
Scalapino, Zhang and Hanke\cite{SCALAPINO1998}. The work of
Schollw\"ock {\it et al} has generated significant interest in
connection with the DDW proposal for the
cuprates\cite{CHAKRAVARTY2001A}.

\begin{figure}
\includegraphics[width=0.4\textwidth]{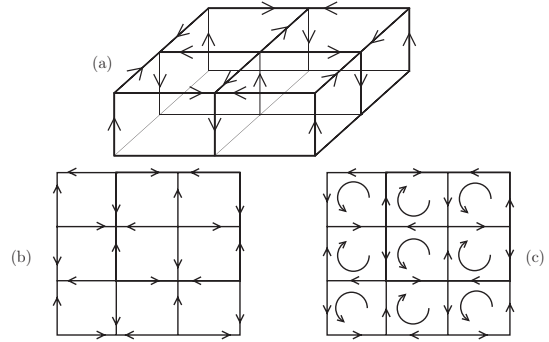}
\caption{\label{fig:current} (a) Sketch of a staggered interlayer
current (SIC) phase. For clarity, we do not show the bottom layer
current. By conservation, each site acts as a source or drain for
the current within the bi-layers. (b) Top view of the the
bi-layer. (c) Sketch of the SF or the DDW current pattern for
comparison.}
\end{figure}

To the best of our knowledge, the existence of a current carrying
ground state has not been conclusively demonstrated in any higher
dimensional models. Following the insights we learned from the 1D
systems, we investigate the current carrying ground state in a
bi-layer version of the model constructed in Ref.
\cite{SCALAPINO1998}. This model was originally constructed and
extensively investigated because of the exact $SO(5)$ symmetry
when coupling constants satisfy a simple relation, and is commonly
referred to as the SZH
model\cite{BOUWKNEGT1999,DUFFY1998,LIN1998,FRAHM2001}. Here we
show that the recently discovered fermionic quantum Monte Carlo
(QMC) algorithm without the  sign problem\cite{WU2003} can also be
applied to this model at and away from half-filling, for a large set of 
parameters, including purely repulsive interactions. Using this
highly accurate numerical method, we can conclusively demonstrate
the existence of a current carrying ground state in this model.
The current carrying ground state is illustrated in Fig.
\ref{fig:current}, with staggered interlayer currents (SIC)
between the bi-layers and alternating source to drain currents
within the bi-layers. Viewed from the top of the bi-layers, this
current pattern is different from the SF or the DDW current
pattern, since it has a $s$-wave symmetry. While the SF or the DDW
currents are divergence free within the layer, the SIC current is
curl free within the layer. These two flow patterns can be
considered as dual to each other in two dimensions. In this paper,
we shall first discuss the physics of the SIC phase by mapping
onto an effective spin one Heisenberg model, and then proceed with
the QMC results.

The Hamiltonian for the SZH model~\cite{SCALAPINO1998} generalized
straightforwardly to the bi-layer system reads
\begin{widetext}
\begin{eqnarray}\label{SZHham} H&=&-t_\parallel \sum_{\langle ij\rangle}
\big\{c^\dagger_{i\sigma} c_{j\sigma} +d^\dagger_{i\sigma}
d_{j,\sigma}+h.c.\big\}-t_\perp \sum_i \Big \{c^\dagger_{i,\sigma}
d_{i,\sigma}+h.c.\Big \} -\mu\sum_i \Big \{c^\dagger_{i,\sigma}
c_{i,\sigma}+ d^\dagger_{i,\sigma} d_{i,\sigma} \Big \}\nonumber
+J\sum_{i}  \vec{S}_{i,c}\cdot
\vec{S}_{i,d}\\
&+& U\sum_{i} (n_{i,\uparrow,c}-1/2)( n_{i,\downarrow,c} -1/2) +
(n_{i,\uparrow,d}-1/2) (n_{i,\downarrow,d}-1/2) +V\sum_{i}
(n_{i,c}-1)(n_{i,d}-1) ,
\end{eqnarray}
\end{widetext}
where $c$ and $d$ denotes fermionic operators in the upper and the
lower layers, respectively, $\sigma$ corresponds to up and down
spins. At half-filling, $\mu=0$, and the model is particle-hole
symmetric. $t_\parallel=1$ sets the unit of energy. Before
discussing the SIC phase, we first discuss some general properties
of the SZH model which were not known before. The SZH model was
known to have a $SO(5)$ symmetry when $J=4(U+V)$ and $\mu=0$, which
unifies antiferromagnetism with
superconductivity\cite{SCALAPINO1998}. Remarkably, it also has
another $SO(5)$ symmetry when
\begin{eqnarray}\label{SO5ph} J=4(U-V),\ \ \ t_\perp=0,
\end{eqnarray}
valid for all filling factors. In order to distinguish between the
two different $SO(5)$ symmetries, we call the former the
particle-particle $SO(5)$ symmetry, denoted by $SO(5)_{pp}$, and
call the latter the particle-hole $SO(5)$ symmetry, denoted by
$SO(5)_{ph}$. The mathematical structure associated with the
$SO(5)_{ph}$ algebra, not necessarily the symmetry itself, plays a
crucial role in constructing the fermionic QMC algorithm without
minus sign problem.

We construct a four component fermion field $\Psi=\{c_\sigma,
d_\sigma\}$. Using the five Dirac $\Gamma_a$ matrices given in
Ref.\cite{WU2003}, we construct the fermion bi-linears \bea
n_a=\Psi^\dagger \frac{\Gamma_a}{2}\Psi \hspace{5mm}
L_{ab}=\Psi^\dagger \frac{\Gamma_{ab}}{2} \Psi \eea It is
straightforward to check that $[H,L_{ab}]=0$ when Eq.
(\ref{SO5ph}) is satisfied, thus demonstrating the exact
$SO(5)_{ph}$ symmetry. The SZH model can be mapped exactly to the
spin $3/2$ Hubbard model\cite{WU2003}, by the identification $
c_\uparrow = c_{3/2},\ c_\downarrow = c_{1/2},\ d_\uparrow =
c_{-1/2},\ d_\downarrow = c_{-3/2}$, and the $SO(5)_{ph}$ symmetry
maps exactly onto the $SO(5)$ symmetry of the spin $3/2$ Hubbard
model. Because of the exact mapping from the SZH model to the spin
$3/2$ Hubbard model, we are able to use the QMC algorithm
discovered in Ref. \cite{WU2003}, which works without the minus
sign problem in a large parameter regime.

\begin{figure}
\centering\epsfig{file=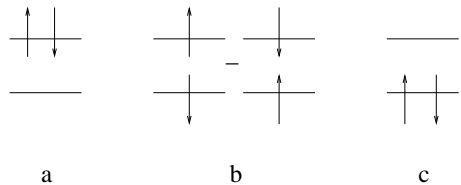,clip=1,width=6cm,angle=0}
\caption{ The double occupancy state a) and c) and the rung
singlet (b). a), b), c) are spin SU(2) singlets and form the
triplet representation of the pseudospin SU(2) group. }
\label{fig:rung}
\end{figure}

In studying the strong coupling phase diagram, SZH identified the
$E_0$ phase where the rung singlet state, depicted in Fig.
\ref{fig:rung}b, is the lowest energy state, and the $E_3$ phase,
where the CDW states, depicted in Fig. \ref{fig:rung}a and
\ref{fig:rung}c are the lowest energy states. The new insight
gained from Ref. \cite{MARSTON2002,SCHOLLWOCK2003} reveals that
the competition between these two phases could result in the DDW
phase. In view of this insight, let us consider the following
operators
\begin{eqnarray*}
n_1(i)&=& i/2\sum_\sigma \big\{c^\dagger_\sigma(i) d_\sigma(i)-
d^\dagger_\sigma(i) c_\sigma(i) \big \}
,\nonumber \\
n_5(i)&=& 1/2 \sum_\sigma \big \{ c^\dagger_\sigma(i) d_\sigma(i)
+d^\dagger_\sigma(i) c_\sigma(i) \big \}, \nonumber \\
Q(i) &=& L_{15}=1/2 \sum_\sigma \big \{ c^\dagger_\sigma(i)
c_\sigma(i) -d^\dagger_\sigma(i) d_\sigma(i) \big \},
\end{eqnarray*}
where $\vec{\sigma}$ are the Pauli matrices. These operators
describe rung current ($n_1$), rung kinetic energy ($n_5$) and the
CDW order parameter ($Q$). These three operators form a
pseudo-spin $SU(2)$ algebra which are important for our discussion
of the SIC phase.

There are 16 states on each rung, including 8 bosonic states with
particle number 0, 2 or 4 and eight fermionic states with particle
number 1 or 3. We are interested in the three rung states shown in
Fig.~\ref{fig:rung}a,b,c, which form a spin-1 representation of
the pseudospin SU(2) algebra defined above, with eigenvalues
$Q=1,0,-1$. $n_1\pm i n_5$ act as pseudospin raising and lowering
operators which connect these three states to each other. At
half-filling and under the condition that $\max(U, V-3/4 J)<
\min(V+J/4, U+2V, U/2+V)$, these are the three lowest energy
states. Furthermore, at $U=V-3/4 J$, these three states are
degenerate. In the strong coupling limit, we can construct an
effective theory to describe the low energy physics by using a
pseudospin-1 antiferromagnetic Heisenberg model
\bea\label{exchange} H_{ex}= J_p \sum_{\langle i,j \rangle} \Big
\{n_5(i) n_5(j) +n_1(i) n_1(j) +Q(i) Q(j) \Big \}, \eea with $J_p=
2 t_\pp^2 / (V+\frac{3}{4} J) $. Several terms break the
pseudospin SU(2) symmetry. The intra-rung hopping $t_\perp$ term
acts as an uniform external magnetic field which couples to $n_5$.
Also, the deviation of $U$ from $V -3/4 J$ removes the degeneracy
between a), c) and b) states. These symmetry breaking terms are
described by the on-site part as \bea\label{break} H_{on}&=&
\sum_{i} \{-2 t_\perp n_5(i) + \Delta U (Q^2(i)-1/2) \} \eea where
$\Delta U= U-(V-3/4 J)$. The nonzero value of $\Delta U$ also
gives different corrections to the three exchange terms  at the
order of $J_p \Delta U /U$. We will neglect these corrections
below because the more important symmetry breaking effect from
$\Delta U$ has already been taking into account in the on-site
part. $H=H_{ex}+H_{on}$ describes a 2D antiferromagnetic spin one
Heisenberg model in an uniform magnetic field $t_\perp$, with
either easy axis ($\Delta U<0$) or easy plane ($\Delta U>0$)
anisotropy.

\begin{figure}
\centering\epsfig{file=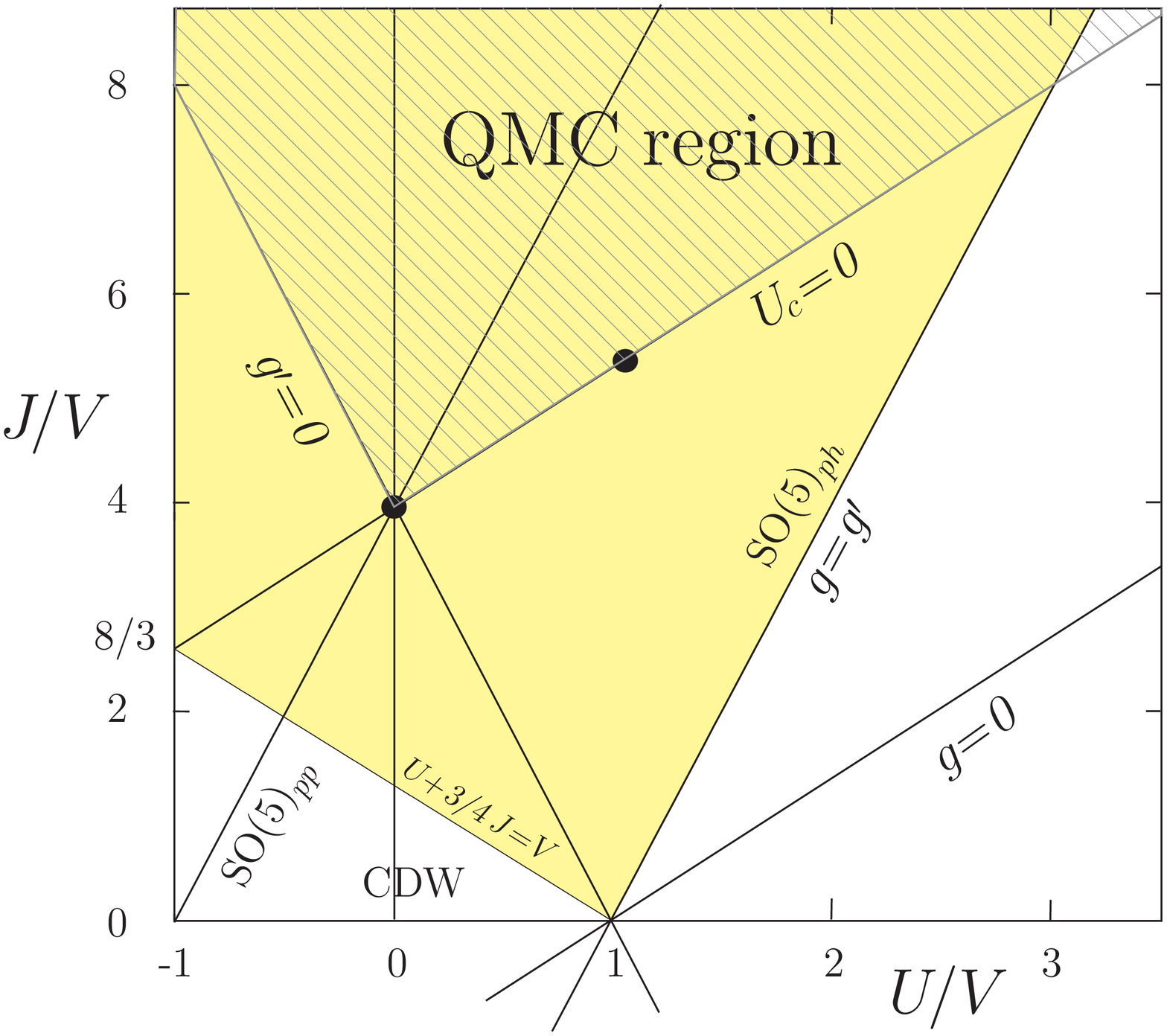,clip=1,width=7cm,angle=0}
\caption{\label{PhaseDiagram} Phase diagram in the strong coupling
limit. Both SO(5) lines are shown as well as QMC region with no
minus-sign problem for any filling (hatched area): $g>0$, $g'>0$
and $U_c>0$. There is also another region with $V<0$ (not shown).
In the yellow region, the low-energy bosonic states are $a$, $b$
and $c$ shown in Fig.~\ref{fig:rung}. This is where we expect the
competition between the SIC and the rung-singlet phase. Black dots
correspond to models for which we have performed QMC simulations.}
\end{figure}

For the easy axis case with $\Delta U<0$, the effective
Hamiltonian reduces to an Ising model with $Q=\pm 1$ states, in a
transverse magnetic field\cite{SCALAPINO1998,TSUCHIIZU2002}. For
$t_\perp =0$, and $\Delta U > 0$, the rung singlet state (b) has
the lowest energy. However, in this case, there is a competition
between the $\Delta U > 0$ term and the Heisenberg exchange term
$J_p$. For $\Delta U
> zJ_p$, where $z=4$ is the coordination number, the ground
state is a featureless Mott insulating state which can be
described as a product of the rung singlet state on each site. On
the other hand, for $\Delta U < zJ_p$, it is more favorable to
form linear combinations between the (a), (b) and (c) states, such
that a staggered ground state expectation value of $\langle n_1
\rangle$ and $\langle n_5 \rangle$ is spontaneously developed,
thus lowering the Heisenberg exchange energy $H_{ex}$. In this
case, and for $t_\perp =0$, the pseudo-spin vector can lie along
in any direction in the $(n_1,n_5)$ plane. At $\Delta U=0$, a
finite value of $t_\perp > 0$ corresponds to a pseudo-spin
magnetic field along the $n_5$ direction, which creates an easy
plane in the $(n_1, Q)$ space. The antiferromagnetic component of
the pseudo-spin moment lies in the $(n_1, Q)$ plane, but the
uniform component of the pseudo-spin moment points along the $n_5$
direction. The pseudo-spin moment becomes fully polarized when
$t_\perp > \frac{z}{2} J_p$, and the antiferromagnetic component
vanishes beyond this point. We see that $t_\perp > 0$ favors the
$(n_1, Q)$ easy plane while $\Delta U < zJ_p$ favors the $(n_1,
n_5)$ easy plane, therefore, when both conditions are satisfied,
the intersection between the two easy planes, namely the $n_1$
easy axis, is selected. This is exactly the staggered inter-layer
current (SIC) order. Combining all these considerations, we can
summarize the subtle criteria for the SIC phase as \bea
&&V-\frac{3}{4} J < U < \min(V+\frac{J}{4}, 2V),~~ V>0 \nonumber \\
&& t_\perp < \frac{1}{2} z J_p \sqrt{1-(\Delta U/z
J_p)^2},~~\Delta U< z J_p, \label{sicMF} \eea The first two robust
conditions ensure that the (a), (b) and (c) states are the lowest
and next lowest energy states among the 16 states on the rung,
while the last two conditions are the rough mean field estimates
discussed above.

On Fig.~\ref{PhaseDiagram}, we show some specific regions on the
phase diagram, obtained in the strong coupling limit.  There are two 
additional axis for $t_\parallel$ and $t_\perp$.  If $t_\parallel$ and/or $t_\perp$ 
gets larger, we can expect some
phases to have larger or smaller extension. In the case of
ladders, a similar phase diagram has been
proposed~\cite{TSUCHIIZU2002,SCHOLLWOCK2003}. In order to obtain
significant current correlations, one should be close enough from
the line $V=U+3/4 J$ shown on Fig.~\ref{PhaseDiagram} where states
a, b and c become degenerate.

Now we proceed to discuss the QMC calculation of the SIC phase. We
first express the interaction terms of the SZH model as
\begin{equation}\label{NLsigma}
 H_{int}=-g (n_1^2 + n_5^2) - g' (n_2^2+n_3^2+n_4^2) - U_c
(n-2)^2,
\end{equation}
up to a constant term. Here $4U_c=-U-3V+3J/4$, $4g=V-U+3J/4$ and
$4g'=U-V+J/4$. The $SO(5)_{ph}$ symmetry is clearly recovered when
$g=g'$, i.e., when $U=V+J/4$. We now introduce auxiliary
Hubbard-Stratonovich fields to decouple each of the three terms
above. Wu, Hu and Zhang\cite{WU2003} have shown that the QMC
algorithm is free of the minus sign problem provided all three
coefficients, $g$, $g'$ and $U_c$ are positive. It corresponds to
a wedge in the phase diagram shown on Fig.~\ref{PhaseDiagram}, and
most remarkably, it includes a region with purely repulsive
interactions, where $U$, $V$ and $J$ are all positive. A simpler
case containing only $n_4^2$ interaction, which explicitly breaks
the SU(2) spin rotation invariance, has been studied in another
context~\cite{ASSAAD2003}. The ground-state (GS) properties of our
model are conveniently studied with the projector auxiliary field
QMC algorithm. The basic idea is to apply the operator $\exp
(-\theta H)$ to a trial state. When $\theta$ becomes large enough
and with a proper normalization, this state converges
exponentially to the GS. Details of the algorithm may be found
in~\cite{ASSAAD1997}. The Trotter discretization was chosen to be
$\Delta \tau=0.1$ but we checked that it does not change the
results.

\begin{figure}
\includegraphics[width=0.8\linewidth]{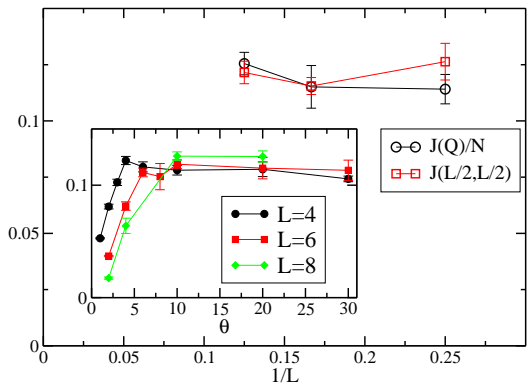}
\caption{\label{model8} Parameters are $t_\perp=0.1$, $U=0$,
$V=0.5$, $J=2.0$ and correspond to $g=0.25$, $g'=0$ and $U_c=0$.
Scaling of ${\cal J}(\vec{Q})/N$ and $J(L/2,L/2)$ vs $1/L$ showing
almost no finite-size effects and proving long-range order in the
thermodynamic limit. (Inset shows the convergence of ${\cal
J}(\vec{Q})/N$ with the projection parameter $\theta$. Typically
the GS value is obtained for $\theta=20$).}
\end{figure}


We compute correlations between rung currents $ n_1(\vec{r})$ and
perform its Fourier transform
\begin{equation}\label{correlator}
{\cal J}(\vec{q})=\frac{1}{N}\sum_{\vec{r}} e^{i \vec{q}\cdot
\vec{r}} \sum_i \langle n_1(i) n_1(i+\vec{r}) \rangle.
\end{equation} The strongest signal
in the Fourier transform is found for $\vec{Q}=(\pi,\pi)$,
suggesting a staggered current pattern as shown in Fig.
\ref{fig:current}. This quantity converges to its GS value as the
projector parameter $\theta$ increases as shown in the inset of
Fig.~\ref{model8}. In order to obtain information in the
thermodynamic limit, one has to make an extrapolation of these GS
values with a $1/L$ finite-size scaling, where $L$ is the linear
size ($L=4$, 6 and 8 in our simulations).

\begin{figure}
\includegraphics[width=0.8\linewidth]{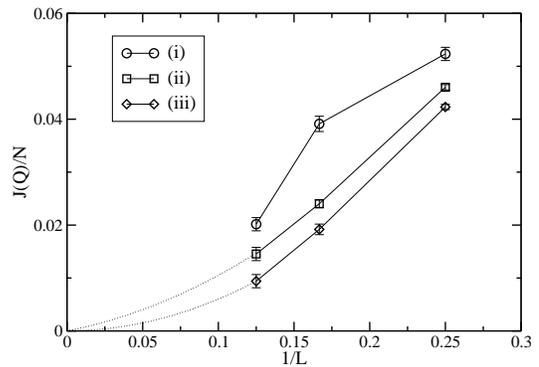}
\caption{\label{noLRO} Finite-size scaling of the current correlations ${\cal
J}(\vec{Q})/N$ showing no long-range order in the thermodynamic limit. Parameters are~:
(i) same as Fig.~\ref{model8} except for $t_\perp=0.5$ (ii) $U=V=0.3$, $J=1.6$ and
$t_\perp=0.5$ at half-filling; (iii) $U=0.75$, $V=0$, $J=1$ and $t_\perp=0$ at 1/8-doping.
 Typically
the GS value is obtained for $\theta=20$.
}
\end{figure}

Following our previous mean-field arguments, in order to prefer a
phase with staggered current, we
choose $g > g'$ and $U_c=0$, with a small $t_\perp$.  As shown on
Fig.~\ref{model8} for $U=0$, $V=0.5$ and $J=2$ when $t_\perp$ is
small at 0.1,  our values are rather constant with size, as expected
in an Ising-like phase. Both the largest distance real-space
correlations $J(L/2,L/2)$ and the Fourier transform ${\cal
J}(\vec{Q})/N$ converge to the same finite value (within our error
bars), meaning long-range order in the thermodynamic limit.

As expected from our analytical estimates in (\ref{sicMF}), if
$\Delta U$ or $t_\perp$ gets too large, long-range order disappear
as shown on Fig.~\ref{noLRO}. Since we can also perform the QMC
simulation at finite-doping without the sign problem, we have
chosen to work at 1/8-doping for some parameters shown on
Fig.~\ref{noLRO}. Again, rung-current correlations vanish in the
thermodynamic limit since the Fermi surface is not nested anymore.
From the analytical estimates based on the mapping to the spin one
antiferromagnetic Heisenberg model and the detailed QMC
calculations shown above, we can conclusively demonstrate the
existence of the SIC phase at half-filling, and also note that
this is a rather subtle phase which can be easily destabilized by
large $U$ and doping.



\acknowledgements This work is supported by the NSF under grant
numbers DMR-0342832 and the US Department of Energy, Office of
Basic Energy Sciences under contract DE-AC03-76SF00515. S.C.
thanks IDRIS (Orsay) and SLAC (Stanford) for allocation of
CPU-time.


\end{document}